\documentclass[preprint,12pt]{elsarticle}
\usepackage[english]{babel}
\usepackage[numbers]{natbib}
\usepackage{hyperref}
\usepackage{lscape}
\usepackage{xcolor}
\usepackage{color}
\usepackage{colortbl}
\usepackage{setspace}
\usepackage{stmaryrd}
\usepackage{geometry}
\usepackage{booktabs}
\usepackage{sidecap}
\usepackage{graphics}
\usepackage{graphicx}  
\usepackage{dsfont}
\usepackage{subcaption}
\usepackage{amsfonts}
\usepackage{amsmath}
\usepackage{amssymb}
\usepackage{amsthm}
\usepackage{bbm}
\usepackage{enumitem}
\usepackage{float}
\usepackage{multirow}
\usepackage{multicol}
\usepackage{array}
\usepackage{adjustbox}
\setcitestyle{square}
\journal{---}
\begin{document}
\begin{frontmatter}

\title{Integrating the implied regularity into implied volatility models: A study on free arbitrage model}
\author[inst1]{Daniele Angelini}
\ead{daniele.angelini@uniroma1.it}

\affiliation[inst1]{organization={MEMOTEF, Sapienza University,  00185, Rome},Department and Organization
            country={Italy}}
\author[inst2]{Fabrizio Di Sciorio}
\ead{fd940@inlumine.ual.es}
\affiliation[inst2]{organization={Department of Economics and Business, University of Almerìa, 04120, Almerìa},
            country={Spain}}




\begin{abstract}

Implied volatility IV is a key metric in financial markets, reflecting market expectations of future price fluctuations. Research has explored IV's relationship with moneyness, focusing on its connection to the implied Hurst exponent $H$. Our study reveals that $H$ approaches $\frac{1}{2}$ when moneyness equals $1$, marking a critical point in market efficiency expectations. We developed an IV model that integrates H to capture these dynamics more effectively. This model considers the interaction between H and the underlying-to-strike price ratio $\frac{S}{K}$, crucial for capturing IV variations based on moneyness. Using Optuna optimization across multiple indexes, the model outperformed SABR and fSABR in accuracy. This approach provides a more detailed representation of market expectations and IV-H dynamics, improving options pricing and volatility forecasting while enhancing theoretical and practical financial analysis.
\end{abstract}

\end{frontmatter}
\section{Introduction}
Since its introduction in the 1970s, the Black-Scholes model has been fundamental to option pricing, linking option prices to implied volatility $\sigma_{BS}$. Initially assuming constant volatility, the model fails to capture empirical complexities, such as skew or smile effects across strike prices and term structures over maturities. Research, including works by Poterba, Summers ~\cite{Poterba1986}, and Stein ~\cite{Stein1989}, introduced refinements like mean-reversion in volatility and sensitivity to short-term shocks. Stochastic volatility models and higher-moment adjustments further extended these insights, with contributions like Corrado \textit{et al.}’s linking volatility skews to skewness and kurtosis ~\cite{corrado1997implied} and Feunou \textit{et al.}’s Homoscedastic Gamma model ~\cite{feunou2017} for more parsimonious skewness handling.

Recent studies focus on the evolving implied volatility surface $\sigma_{BS}\left(K, \tau,t\right)$, influenced by strike, maturity, and market conditions. This is particularly significant for currency options, as highlighted by Chalamandaris and Guo ~\cite{Chalamandaris2010b, Guo2014}. Related fields examine VIX term structures and realized volatility as predictors of market behavior. Advanced models, including rough volatility frameworks and fractional extensions like the Fractional Black-Scholes Inspired model ~\cite{flint2017}, address memory effects and short-maturity dynamics using fractional Brownian motion. Approximations like the ADO-Heston model streamline computations while retaining rough volatility features, tackling at-the-money skews and VIX dynamics ~\cite{forde2017}.
Recent approaches, including Bianchi \textit{et al.}'s ~\cite{bianchi2022forecasting} use of the Hurst exponent and multifractional Brownian motion, enhance VIX forecasting and risk management during market turbulence.

Building on these advances, this paper introduces a model linking implied volatility to the Hurst exponent $H$ and moneyness, uncovering an inverse smile effect: $H$ peaks at-the-money (ATM) and decreases in in-the-money (ITM) and out-of-the-money (OTM) regions. Integrated into the Black-Scholes framework under no-arbitrage conditions, this model offers deeper insights into implied volatility’s structural properties and smoothness, addressing limitations of traditional approaches while enhancing the continuity analysis of the volatility surface. The paper is structured as follows: Sections \ref{fbm_FMH} and \ref{fSABR} briefly introduce fractional markets and the Hurst exponent, discussing fractional stochastic processes and their role in financial modeling. Section \ref{closed_formula} presents a closed-form model linking implied volatility to the Hurst exponent and moneyness, explaining its predictive mechanisms. Section \ref{application} applies the model to real data, comparing it with SABR and fractional SABR models. Section \ref{conclusion} concludes.
\subsection{Fractional Brownian motion and Fractional Market Hypothesis} \label{fbm_FMH}

Fractional Brownian motion (fBm), introduced by Mandelbrot and Van Ness ~\cite{mandelbrot1968}, provides a more realistic model for asset dynamics compared to standard Brownian motion and effectively captures key stylized facts of financial time series. Widely used in studying market phenomena with long-term dependencies, fBm is characterized by the Hurst exponent $H\in(0,1)$, which quantifies the rate of decay of autocorrelation. For $H>\frac{1}{2}$, the series exhibits persistence, continuing in the same direction; for $H<\frac{1}{2}$, it shows anti-persistence, with a tendency for reversals. When $H=\frac{1}{2}$, the series behaves as a continuous random walk, consistent with Brownian motion. The fBm process is defined as follows ~\cite{couillard2005}
\begin{equation}\label{eq:fbm}
    B_t^H = \frac{k\sqrt{\Gamma(2H+1)\sin(\pi H)}}{\Gamma\left(H + \frac{1}{2}\right)} \int_{-\infty}^{t}\left[(t-\tau)_+^{H-\frac{1}{2}} - (-\tau)_+^{H-\frac{1}{2}}\right] dW_{\tau},
\end{equation}
where $k$ is a scale parameter and $dW_t$ denotes Brownian measure. The covariance structure of the fBm process is given by
\begin{equation}
    \mathbb{E}[B_t^H B_s^H] = \frac{k^2}{2} \left( |t|^{2H} + |s|^{2H} - |t - s|^{2H} \right),
\end{equation}
where $k$ is such that $k^2 = \mathbb{V}ar[B_1^H]$.
\noindent
The \textit{Fractal Market Hypothesis} (FMH), introduced by Peters ~\cite{peters1994fractal} and inspired by Mandelbrot, explains turbulence and instability in financial markets. It emphasizes the importance of market liquidity and varied investment horizons for stability. When all participants interpret information similarly, liquidity decreases, leading to instability and potential market collapses, especially under short-term trader dominance.

FBm, central to FMH, has been increasingly applied to model rough volatility. Studies, including Gloter \textit{et al.} ~\cite{gloter2004}, highlight the roughness of market volatility, often associated with a Hurst parameter below $\frac{1}{2}$ ~\cite{neuman2018}, reflecting fBm’s capability to model fractal structures and long-range dependence. Integrating fBm into volatility models has improved understanding of persistent and erratic market behaviors. The next section introduces fSABR, a fractional stochastic model for implied volatility.
\subsection{Fractional SABR model} \label{fSABR}
The log-normal fSABR model ~\cite{kim2020volatility,akahori2022} is a variation of the standard SABR model ~\cite{hagan2015} that incorporates fBm into the stochastic process governing volatility. This model provides a more flexible framework for capturing market dynamics, especially in the presence of long-range dependence in volatility. In the log-normal fSABR model, the underlying asset price dynamics are given by
\begin{equation}\label{eq:fSABR}
\begin{cases}
    \frac{dS_t}{S_t} = \alpha_t (\rho dW_{1,t} + \sqrt{1-\rho^2} dW_{2,t}),\\
    \alpha_t = \alpha_0 e^{\nu B_t^H},
\end{cases}
\end{equation}
where $S_t$ represents the asset price, $W_{1,t}$ and $W_{2,t}$ are independent Brownian motions, $\alpha_t$ is the stochastic volatility process, $B_t^H$ is an fBm defined as in equation \eqref{eq:fbm} driven by $W_{1,t}$ and $\rho$ is the correlation between the Brownian motion of the asset price $S_t$, which is given by $W_t = \rho W_{1,t} + \sqrt{1-\rho^2}W_{2,t}$ and the fBm $B_t^H$. In the volatility process $\alpha_t$ the parameter $\nu$ control the volatility of volatility.\\
The goal of this model is to obtain an easily accessible expression for the joint density of $(S_t, \alpha_t)$, which is crucial for pricing options and derivatives in markets exhibiting volatility dynamics characterized by long-range dependence.

The main problem of SABR ~\cite{Lund2023} and fSABR models suffer from intrinsic limitations that make them less effective at representing the extreme curvatures of the volatility smiles in ITM and OTM options. Their static assumptions ~\cite{ lesniewski2014}, such as constant correlation $\rho$ and a less dynamic volatility of volatility $\nu$, constrain their ability to adapt to the complex changes in volatility behavior in extreme moneyness regions.

\section{Angelini - di Sciorio closed formula} \label{closed_formula}

\noindent
The closed formula proposed in the paper, termed the Angelini-di Sciorio model (AdS), expresses the dependency of implied volatility $\sigma$ on the moneyness, i.e. ratio between the underlying asset price $S$ and the strike price $K$ of the option, as known in the literature ~\cite{kachhara2022option}. Additionally, the model integrates the effect of memory, represented by the Hurst exponent $H$, along with the sensitivity of volatility to moneyness.

The formula given by
\begin{equation}\label{eq:AdS model 1}
    \sigma\left(\frac{S}{K}\right) = \alpha \left(\frac{S}{K} - \frac{S}{K_{\min}}\right)^2 e^{-\beta H\left(\frac{S}{K}\right) \left(\frac{S}{K} - \frac{S}{K_{\text{min}}}\right)} + \epsilon
\end{equation}
is structured by a quadratic term $\left(\frac{S}{K}-\frac{S}{K_{\min}}\right)^2$, that from a financial perspective represents how much "in or out of the money'' an option is ($K_{\min}$ is the strike price at which $\sigma$ attains its minimum value), and by an exponential term $e^{-\beta H\left(\frac{S}{K}\right)\left(\frac{S}{K}-\frac{S}{K_{\min}}\right)}$ that models how implied volatility decays as the moneyness increases. The former term captures the typical concave behavior of volatility surfaces, which is particularly noticeable in the ITM and OTM regimes, and it is modulated by the $\alpha$ coefficient which adjusts the overall magnitude of the volatility. The latter term is influenced by the Hurst exponent $H$ where $\beta$ coefficient controls the influence of $H$ on the volatility decay: if the memory is strong, volatility decays more slowly, while the memory is weak, volatility decays more quickly. Finally $\epsilon$ term accounts for the small fluctuations and imperfections that are inherent in real-world financial markets, ensuring that volatility never reaches zero.  

The second key element is the function  $H(\frac{S}{K})$ defined as
\begin{equation}\label{eq:AdS model 2}
H\left(\frac{S}{K}\right) = \frac{1}{2} \left( 1 + \left| 1 - \frac{S}{K_{\min}}\right|^\delta \right) \frac{1}{\left( 1 + \left| \frac{S}{K} - \frac{S}{K_{\min}} \right|^\delta \right)}.
\end{equation}

\begin{figure}[!ht]
  \centering
  \begin{subfigure}[b]{0.45\textwidth}
    \centering
    \includegraphics[width=\textwidth]{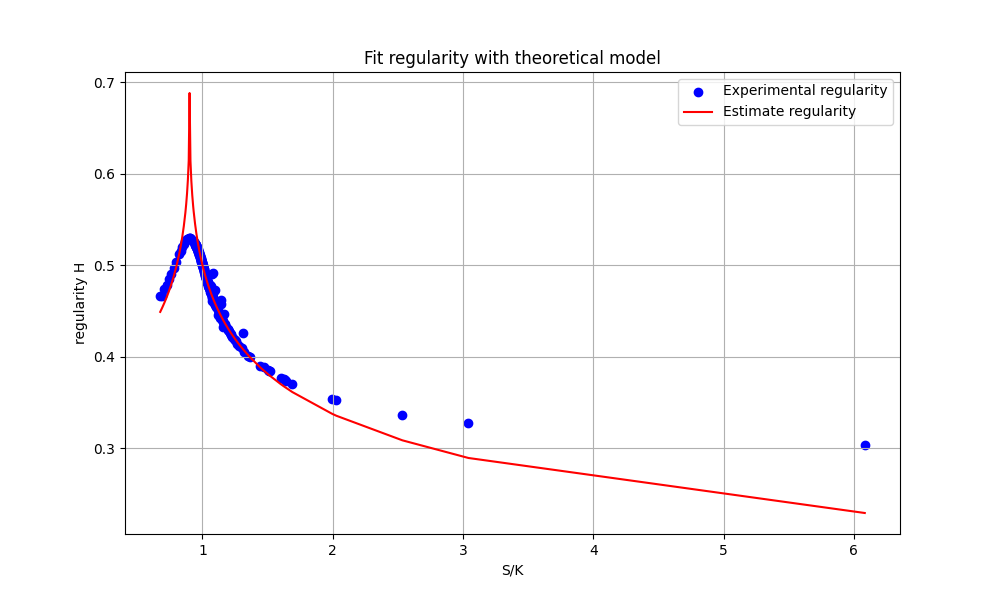}
    \caption{\small Implied regularity (SPX)}
    \label{fig: Implied regularity SPX}
  \end{subfigure}
  \hfill
  \begin{subfigure}[b]{0.45\textwidth}
    \centering
    \includegraphics[width=\textwidth]{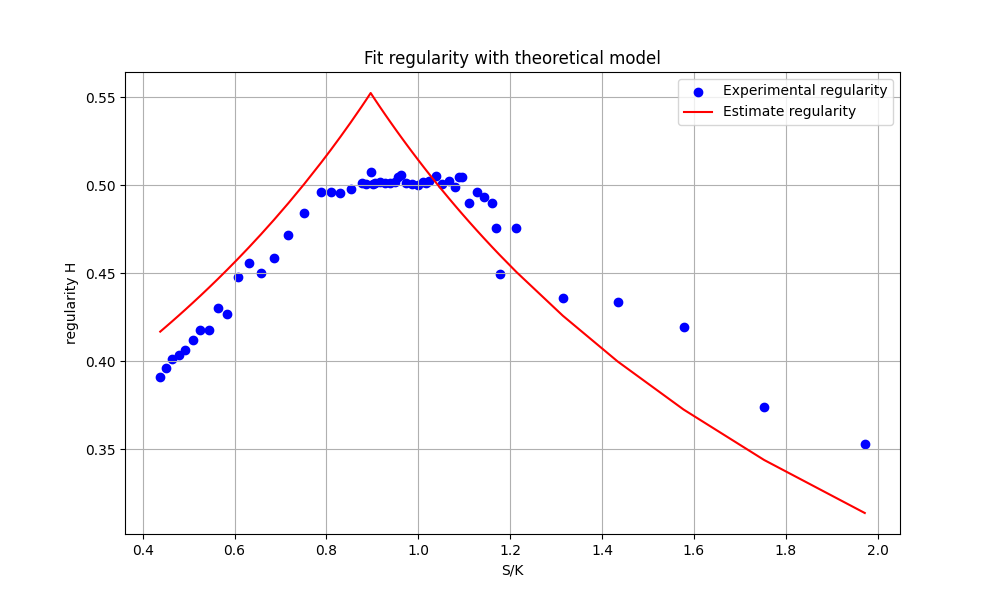}
    \caption{\small Implied regularity (NKE)}
    \label{fig:Implied regularity NKE}
  \end{subfigure}
  \vspace{0.5cm}
  \begin{subfigure}[b]{0.45\textwidth}
    \centering
    \includegraphics[width=\textwidth]{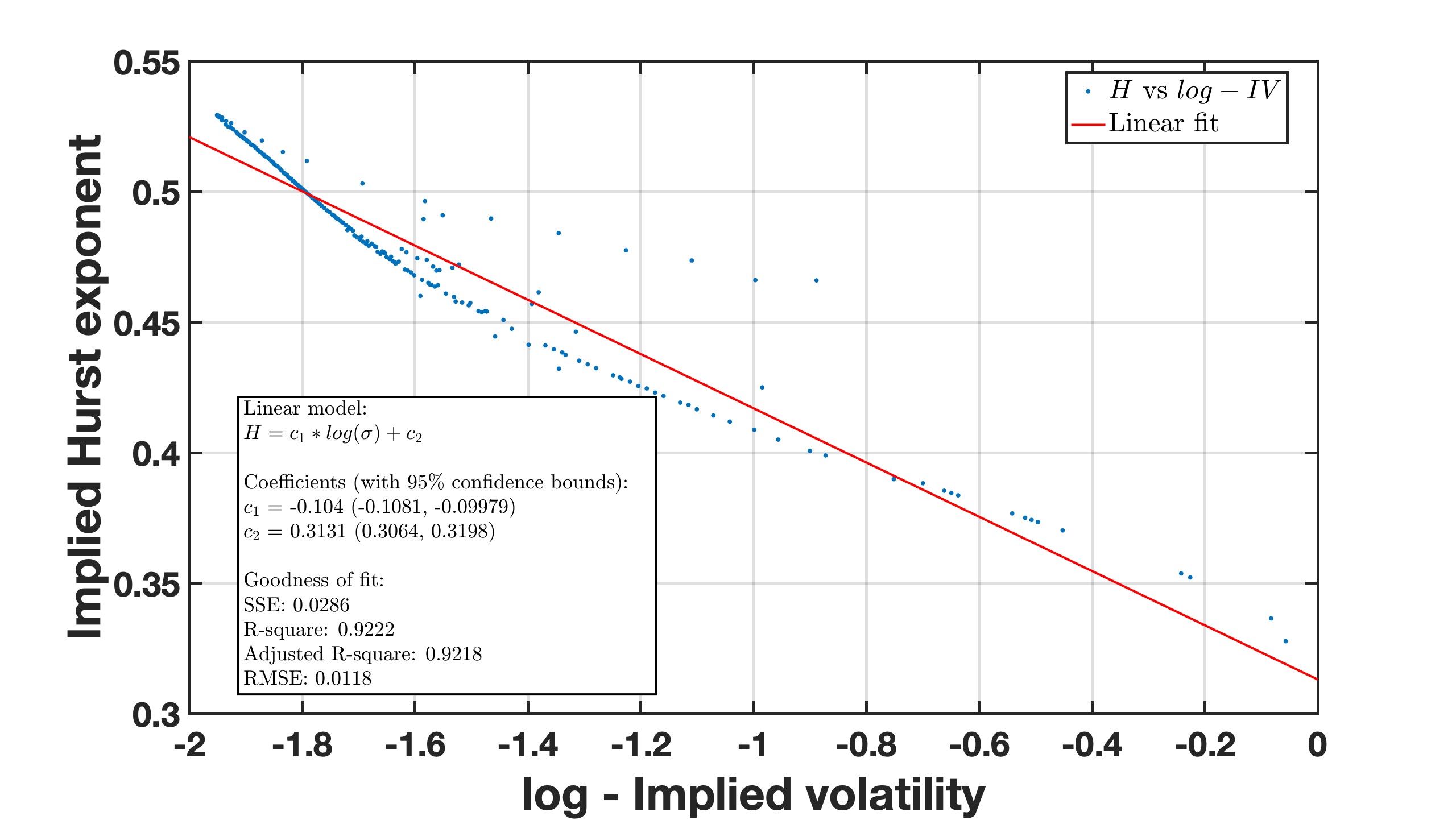}
    \caption{\small Implied regularity vs IV}
    \label{fig:H vs IV SP500}
  \end{subfigure}
  \hfill
  \begin{subfigure}[b]{0.45\textwidth}
    \centering
    \includegraphics[width=\textwidth]{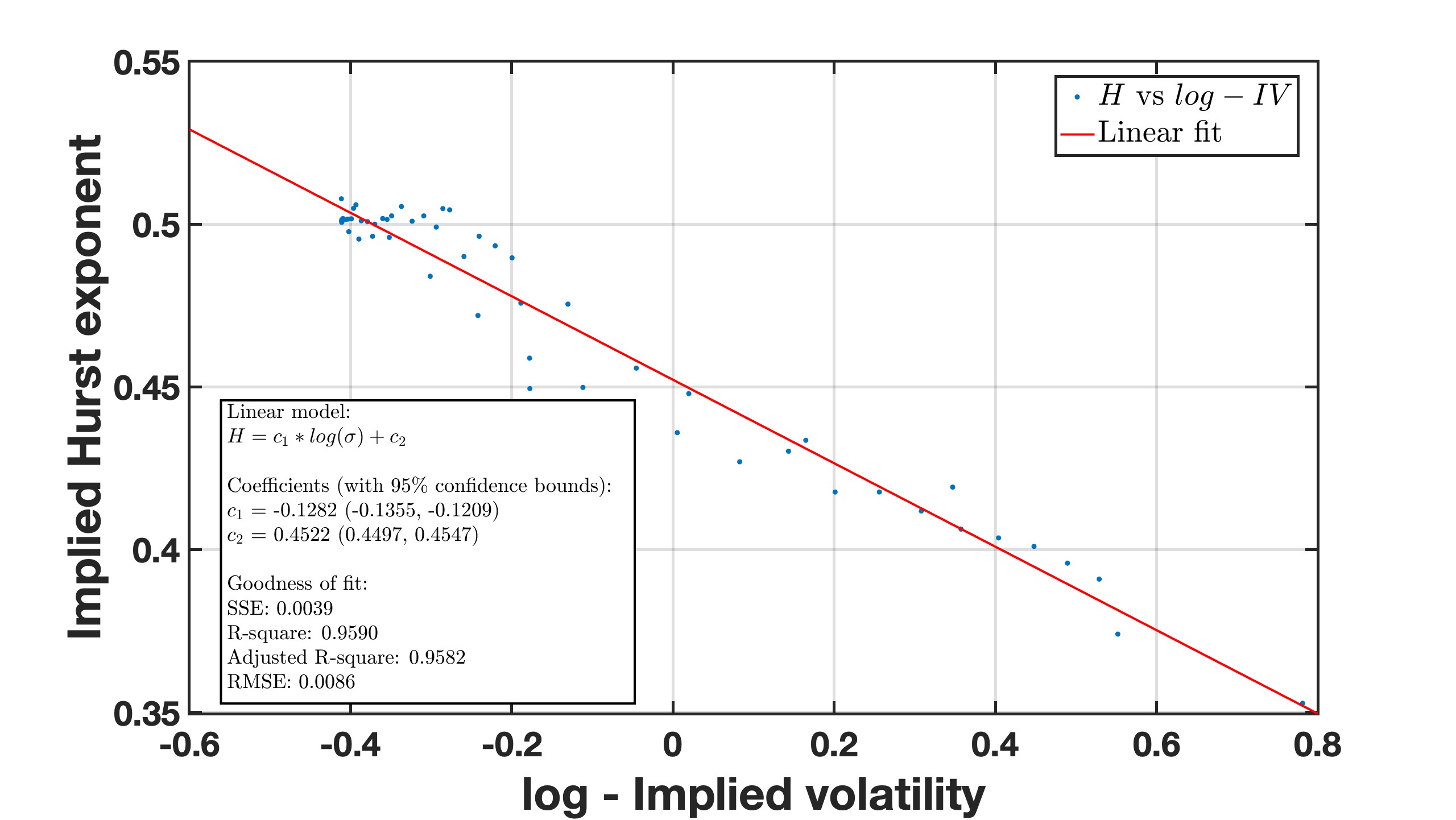}
    \caption{\small Implied regularity vs IV}
    \label{fig:H vs IV NKE}
  \end{subfigure}
  \caption{\small Implied regularity, IV, Moneyness relationship.}
  \label{fig:boxplot_error_metrics 1}
\end{figure}

$H$ depends on the parameter $\delta$ which controls the shape and steepness of the function. The term $\frac{1}{2}\left(1+\left\vert 1-\frac{S}{K_{\min}}\right\vert^\delta\right)$ normalizes $H$ in the region $(0,1)$ and forces it to $\frac{1}{2}$ when $\frac{S}{K}=1$. Essentially, this function determines how long-term memory (represented by $H$) changes based on the moneyness of the option. A distinctive feature of our model lies in the observed behavior of $H$, which exhibits an \textit{inverse smile effect}. Specifically, the Hurst value is $\frac{1}{2}$ in the ATM region, and it decreases as we move towards the ITM and OTM areas. This phenomenon suggests that when the option is in the ATM regions, the price dynamics tend to align with the \textit{Efficient Market Hypothesis} (EMH) framework, as fBm reduces to geometric Brownian motion, as observed in the literature by ~\cite{flint2017}. By contrast, it becomes less relevant in the ITM and OTM regions, where volatility is more reactive to market movements. Therefore, when the option is ATM, the market is informationally efficient, and the option's value primarily depends on the temporal component.
\noindent
In Figure \ref{fig:boxplot_error_metrics 1} we report the empirical and theoretical behavior of $H$ or $implied$ $regularity$ observed on the SPX. The implied regularity has been estimated as in Angelini \textit{et al.} ~\cite{angelini2023}. The authors derive the relationship between $\sigma$ and $H$ starting from the self-similarity property of fBm. It follows that for the discretized sample 
$X_j^H$ ($j \in \llbracket1, n\rrbracket$) of fBm over the interval $t \in [0, 1]$, with 
$t = \frac{j-1}{n-1}$, one has
\[
\sigma^2(n) = \mathrm{Var}(X_{j+1}^H - X_j^H) = C^2 n^{-2H},
\]
which can be linearized as
\begin{equation}\label{eq: H-ss}
\log \sigma(n) = \log C - H \log n.
\end{equation}
The unknown scale parameter $C$ was estimated following the Bianchi \textit{et al.}'s approach ~\cite{bianchi2013}.  The relationship between implied regularity and moneyness is stable regardless of the estimation method used for $H$ (self-similarity, multiscaling, AMBE). 

In addition, we have verified that the Black\&Scholes framework integrated with the AdS model satisfies all five properties of an arbitrage-free volatility surface, as outlined by Zaugg \textit{et al.} ~\cite{zaugg2024}. Specifically, the authors propose a set of conditions (5) for options to ensure that they are free from arbitrage opportunities, including conditions on convexity, behavior at the limits, and the consistency of pricing functions. These conditions help guarantee that the pricing surfaces for both call and put options are arbitrage-free, meaning that no riskless profit opportunities can be exploited from the prices derived from the volatility surface. The detailed verification of these conditions is provided in Section \ref{sec:Annex}. 

\subsection{Calibration}
Calibrating stochastic volatility models ensures alignment between theoretical models and market data by minimizing discrepancies in option prices. Techniques such as Genetic Algorithm ~\cite{Goldberg1989}, Simulated Annealing ~\cite{kirkpatrick1983optimization}, Gauss-Newton with Trust Region ~\cite{marquardt1963algorithm}, and Generalized Reduced Gradient Method ~\cite{abadie1970application}, address challenges like non-convexity and multiple local minima.

Deep learning has recently emerged as a robust solution for calibrating complex stochastic volatility models ~\cite{yuan2024deep, Horvath2021}, particularly in rough volatility or non-Markovian contexts. It employs a two-step approach: training a neural network on synthetic data for fast pricing approximations and using traditional optimization techniques for efficient parameter calibration.

For our model, we utilized the Optuna algorithm\footnote{https://optuna.org/ : function \textit{create\_study(direction = ``minimize")}.} ~\cite{optuna2019}, known for its efficient and automated optimization. Optuna combines Bayesian Optimization and Tree-Structured Parzen Estimator methods to minimize the objective function, in this case, root mean squared error (RMSE). By sampling hyperparameters within a defined search space and constructing a probabilistic model of the objective function, Optuna effectively balances exploration of new values with exploitation of known promising areas, ensuring efficient optimization.

The search space for the model parameters $\alpha$, $\beta$, $\delta$, $\epsilon$ is defined as follows:
\begin{equation}
\mathcal{D} = \left\{ (\alpha, \beta, \delta, \epsilon) \in \mathbb{R}^4 : \alpha >0, \, \beta \in [-1,1], \, \delta \in [0, 1], \epsilon \ >0
\right\}.
\end{equation}
\noindent
Each parameter domain is compatible with the no-arbitrage condition of the model, see \ref{theorem: arbitrage}.

\subsection{Empirical Analysis} \label{application}
\noindent
We applied the model to market data from Yahoo Finance using the Python library  \textit{yfinance}\footnote{https://pypi.org/project/yfinance/ : function \textit{Ticker()}.}, analyzing historical data for U.S. market indices and stocks. We retrieved 30-day expiration options, including strike prices, implied volatilities, and closing prices.

\begin{figure}[!ht]
  \centering
  \begin{subfigure}[b]{0.40\textwidth}
    \centering
    \includegraphics[width=\textwidth]{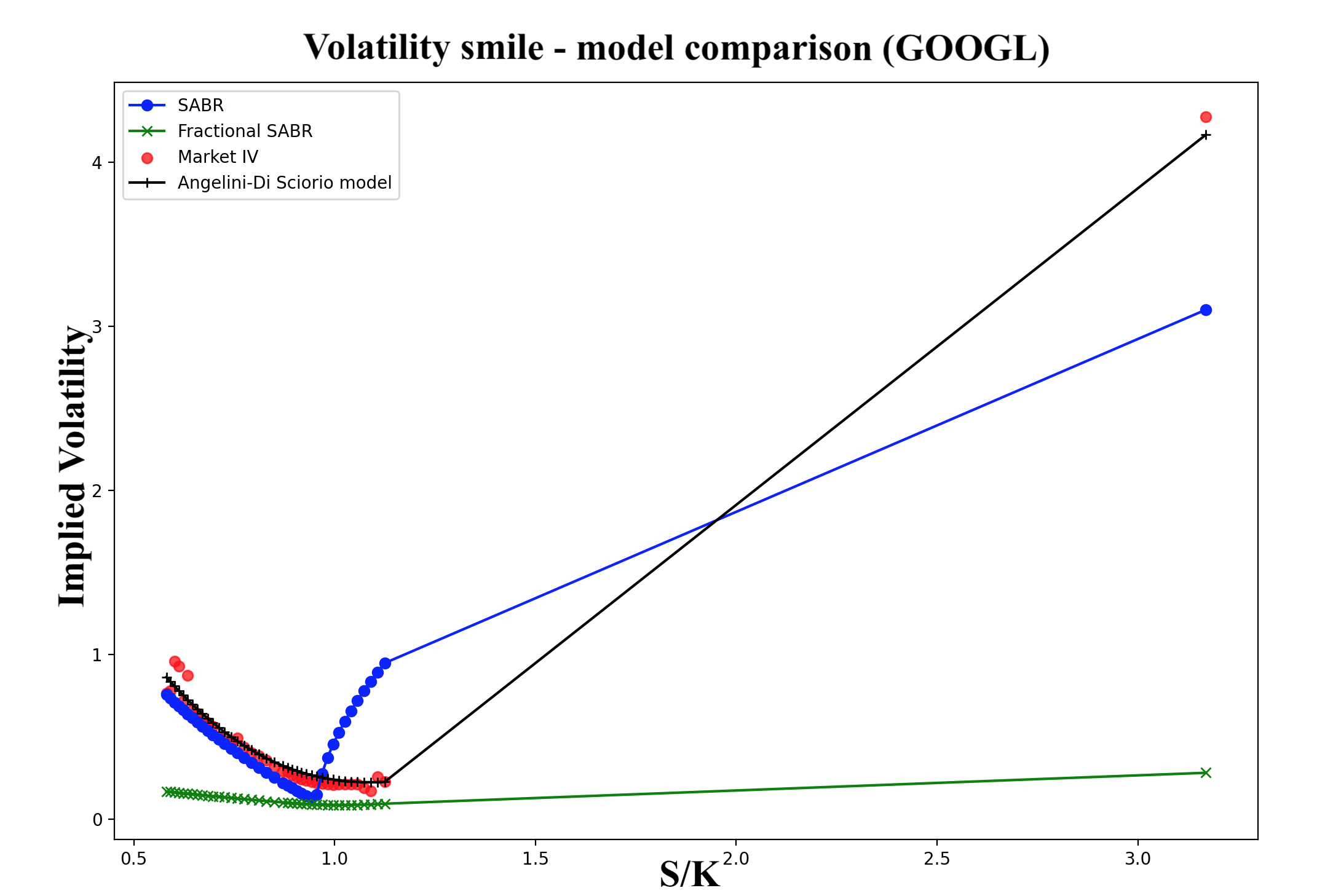}
    \caption{\small IV (GOOGLE)}
    \label{fig:Iv_GOOGLE}
  \end{subfigure}
  \begin{subfigure}[b]{0.45\textwidth}
    \centering
    \includegraphics[width=\textwidth]{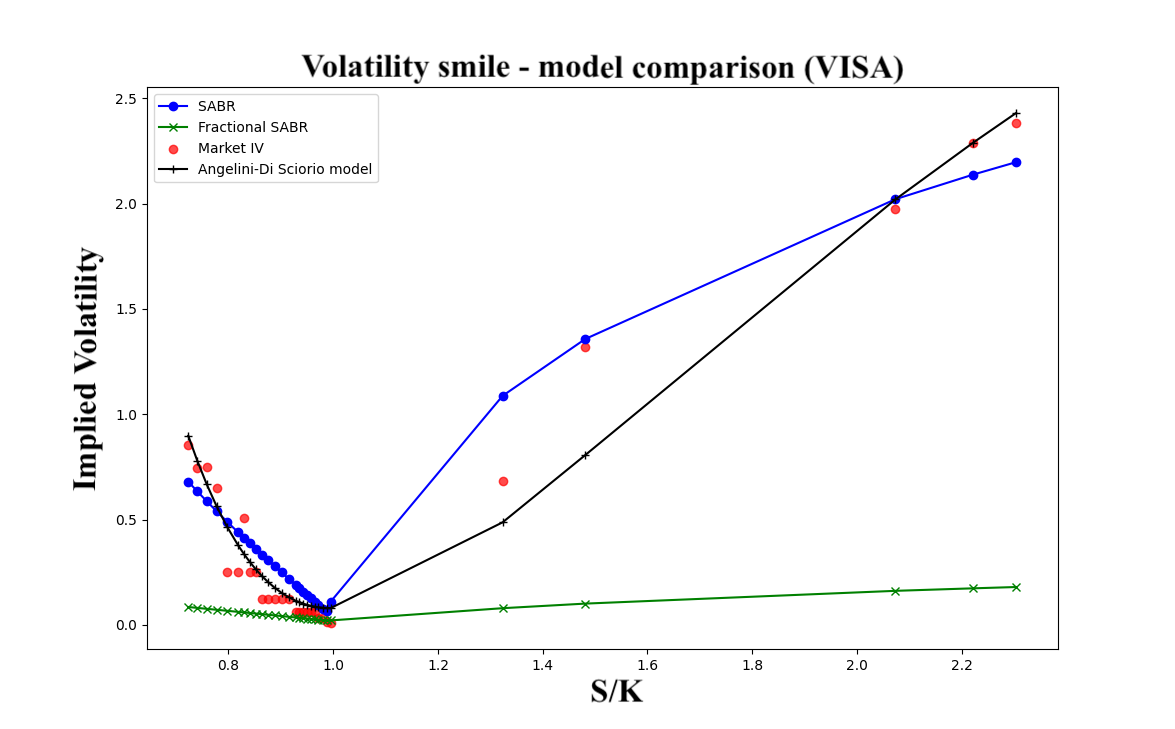}
    \caption{\small IV (VISA)}
    \label{fig:Iv_VISA}
  \end{subfigure}
  \\
  \begin{subfigure}[b]{0.45\textwidth}
    \centering
    \includegraphics[width=\textwidth]{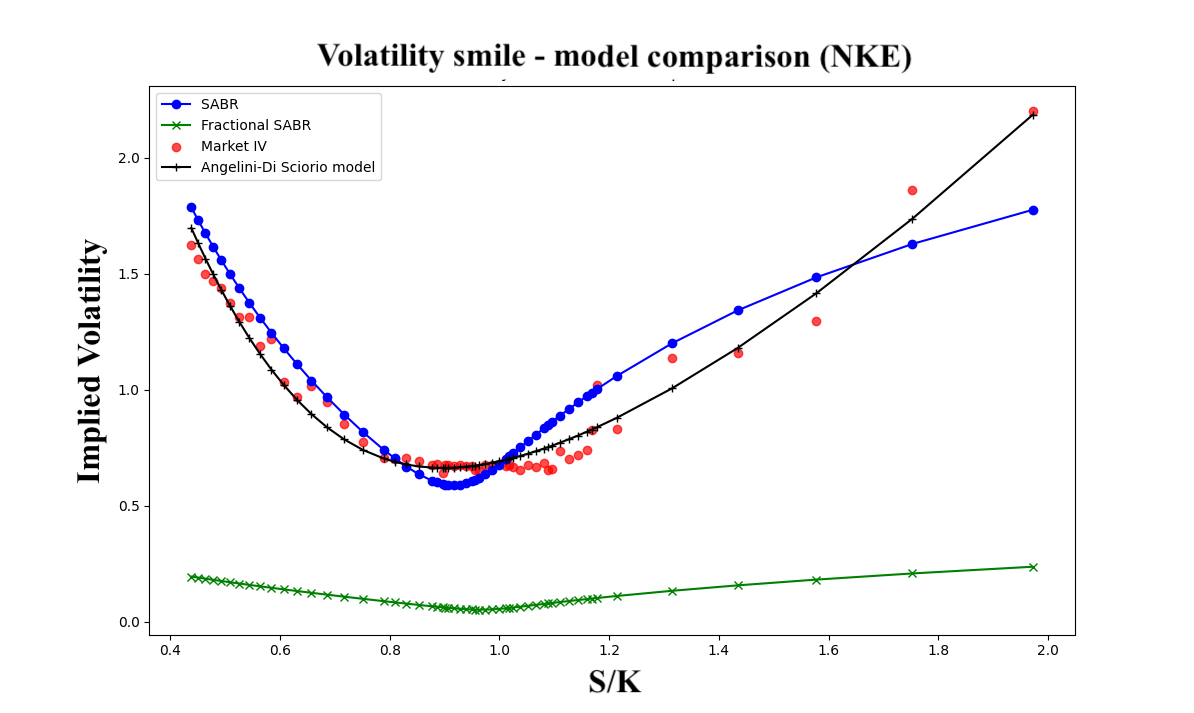}
    \caption{\small IV (NKE)}
    \label{fig:Iv_JNJ}
  \end{subfigure}
  \begin{subfigure}[b]{0.45\textwidth}
    \centering
    \includegraphics[width=\textwidth]{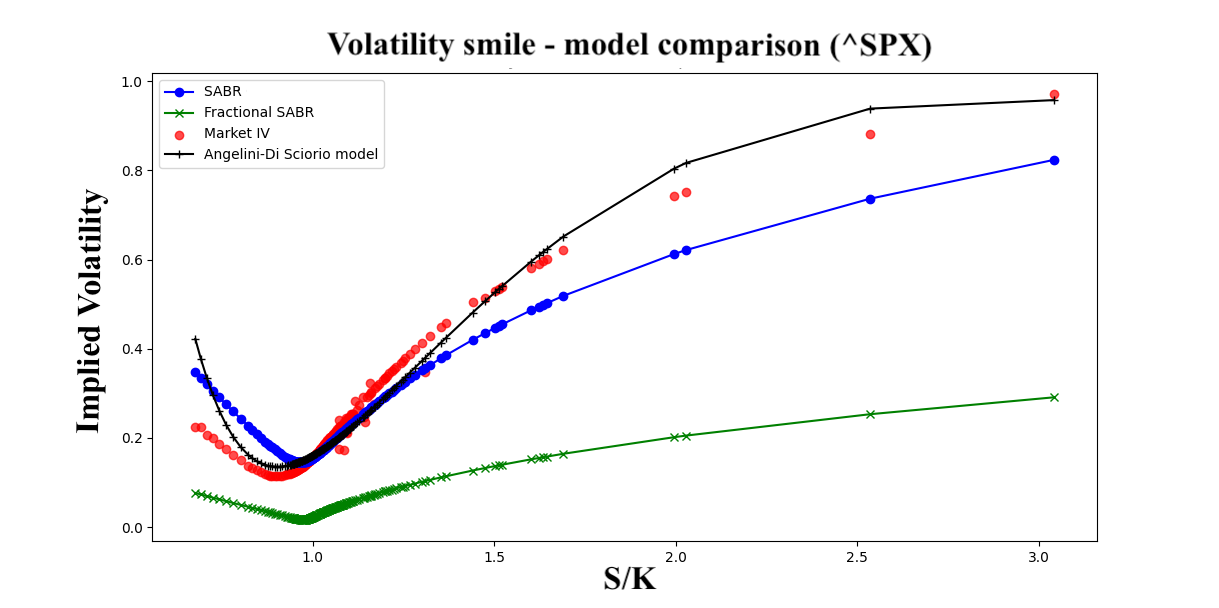}
    \caption{\small IV (SPX)}
    \label{fig:Iv_SPX}
  \end{subfigure}
  \caption{\small Implied Volatility (IV) fitting using SABR, fSABR, and AdS models.}
  \label{fig:IV_fitting}
\end{figure}

\noindent
We compare the performance of the AdS model against the SABR model and its fractional extension, the fSABR model. To ensure a valid and unbiased comparison among the models, we calibrated all models using Optuna, with 100 trials, and minimizing RMSE.
We introduce mean square error (MSE), mean absolute error (MAE) and the curvature error metrics $C$ of IV based on the second derivative of implied volatility $\sigma$ with respect to the moneyness $M$. Formally, the latter is expressed as:
\begin{equation}
  C = \frac{\partial^2 \sigma}{\partial M^2}.  
\end{equation}
Since market data and models provide discrete values for $M$ and $\sigma$, the second derivative is approximated numerically using the central finite difference method. Let us assume a discrete dataset $\{(M_i, \sigma_i)\}_{i=1}^N$, where $M_i$ represents the available moneyness and $\sigma_i$ represents the observed or modeled implied volatilities. The discrete approximation of the second derivative at an internal point $i\in\llbracket 2,N-1\rrbracket$ is given by:
\begin{equation}
   C_i = \frac{\sigma_{i+1} - 2\sigma_i + \sigma_{i-1}}{(M_{i+1} - M_i)^2}. 
\end{equation}
Once the curvature is computed for both market data $C_{\text{obs}}$ and model predictions $C_{\text{mod}}$, we can define error metrics: ACE (absolute curvature error)  and RMSCE (root mean square curvature error) to evaluate the model's performance.
This metric measures the mean absolute error between the modeled and observed curvatures:
\begin{equation}
    ACE = \frac{1}{N} \sum_{i=1}^N |C_{\text{mod}, i} - C_{\text{obs}, i}|,
\end{equation}
where $C_{\text{mod}, i}$ and $C_{\text{obs}, i}$ are the curvatures computed from the model and the curvatures observed in market data at moneyness $M_i$, respectively.
To penalize larger errors more heavily, we can use a quadratic metric
\begin{equation}
    RMSCE = \sqrt{\frac{1}{N} \sum_{i=1}^N (C_{\text{mod}, i} - C_{\text{obs}, i})^2}.
\end{equation}
This metric is sensitive to large deviations between $C_{\text{mod}}$ and $C_{\text{obs}}$, making it particularly useful for diagnosing models that exhibit systematic errors in the extreme regions of the volatility smile.    
Table \ref{tab:1} reports the summary results, the whole dataset is in \ref{sec:data}.

\begin{table}[!h]
\centering
\resizebox{\textwidth}{!}{
\begin{tabular}{|l|c|c|c|c|c|c|c|c|c|c|c|c|}
\toprule
Statistic & AdS\_MSE & AdS\_MAE & AdS\_RMSCE & AdS\_ACE & SABR\_MSE & SABR\_MAE & SABR\_RMSCE & SABR\_ACE & fSABR\_MSE & fSABR\_MAE & fSABR\_RMSCE & fSABR\_ACE \\
\midrule
mean      & 0.503844  & 0.162722  & 0.097482  & 0.036134  & 0.678031  & 0.209677  & 0.101762  & 0.047981  & 0.658167  & 0.210455  & 0.101156  & 0.048044  \\
std       & 2.237659  & 0.389505  & 0.247270  & 0.101942  & 2.932874  & 0.446355  & 0.235415  & 0.125170  & 2.860042  & 0.446364  & 0.236328  & 0.125439  \\
min       & 0.000400  & 0.014900  & 0.000540  & 0.000237  & 0.000600  & 0.019700  & 0.000551  & 0.000238  & 0.000700  & 0.019700  & 0.000559  & 0.000239  \\
25\%      & 0.003177  & 0.040140  & 0.003897  & 0.002135  & 0.007175  & 0.058250  & 0.003822  & 0.002131  & 0.007850  & 0.058425  & 0.003850  & 0.002144  \\
50\%      & 0.008248  & 0.059200  & 0.010790  & 0.005591  & 0.014000  & 0.102200  & 0.009819  & 0.005785  & 0.015450  & 0.102400  & 0.010064  & 0.005848  \\
75\%      & 0.035425  & 0.117350  & 0.054350  & 0.019328  & 0.050950  & 0.158200  & 0.090625  & 0.027100  & 0.042450  & 0.155300  & 0.087125  & 0.027350  \\
max       & 12.725600 & 2.042900  & 1.220200  & 0.626500  & 16.264000 & 2.287000  & 1.260700  & 0.642700  & 16.023000 & 2.288900  & 1.272200  & 0.645000  \\
\bottomrule
\end{tabular}
}
\caption{ }
\label{tab:1}
\end{table}


\begin{figure}[!ht]
  \centering
  \begin{subfigure}[b]{0.35\textwidth}
    \centering
    \includegraphics[width=\textwidth]{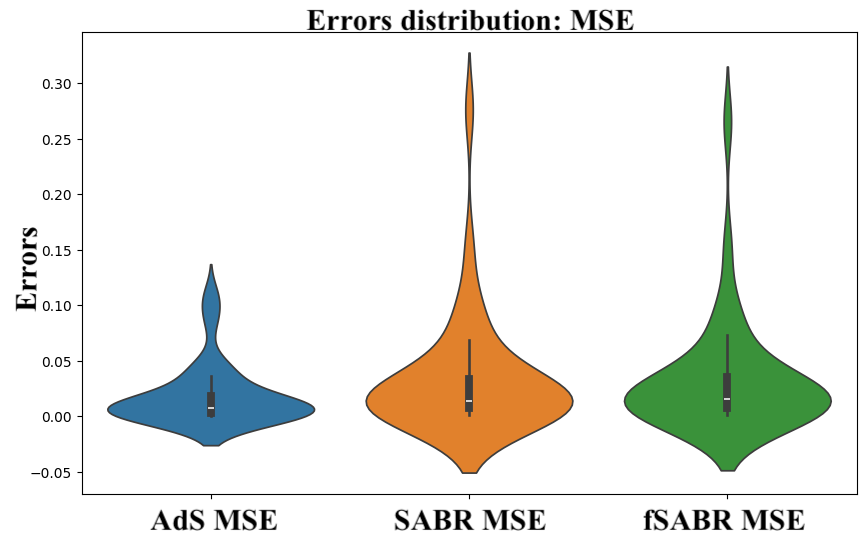}
    \caption{\small Violin plot MSE}
    \label{fig:MSE_errors}
  \end{subfigure}
  \hfill
  \begin{subfigure}[b]{0.35\textwidth}
    \centering
    \includegraphics[width=\textwidth]{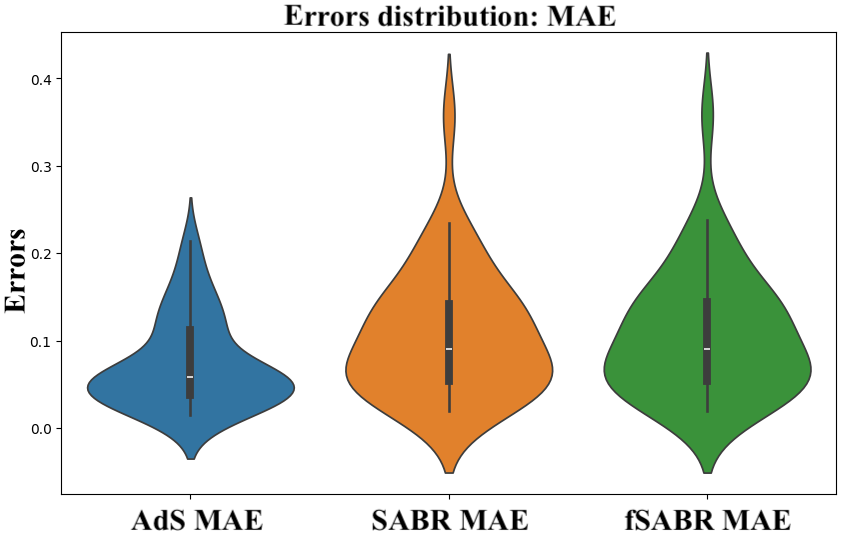}
    \caption{\small Violin plot MAE}
    \label{fig:MAE_errors}
  \end{subfigure}
  \vspace{0.5cm}
  \begin{subfigure}[b]{0.35\textwidth}
    \centering
    \includegraphics[width=\textwidth]{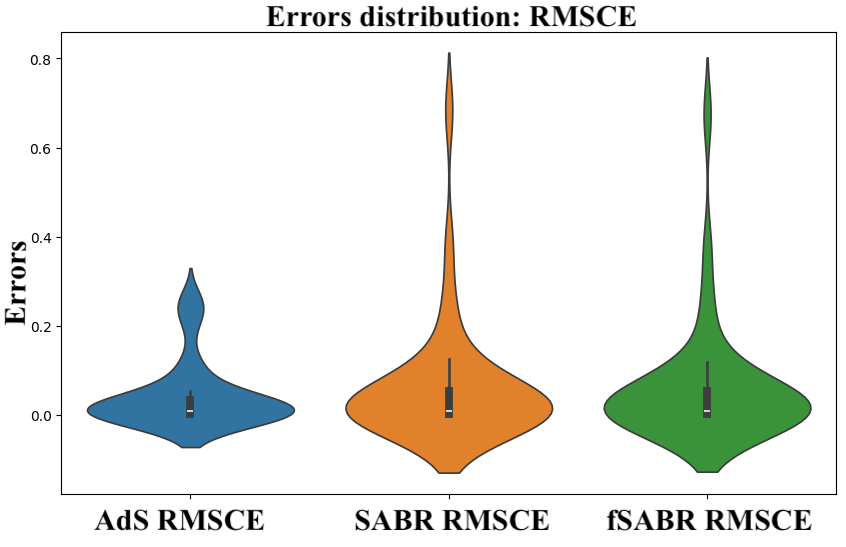}
    \caption{\small Violin plot RMSCE}
    \label{fig:RMSCE_errors}
  \end{subfigure}
  \hfill
  \begin{subfigure}[b]{0.35\textwidth}
    \centering
    \includegraphics[width=\textwidth]{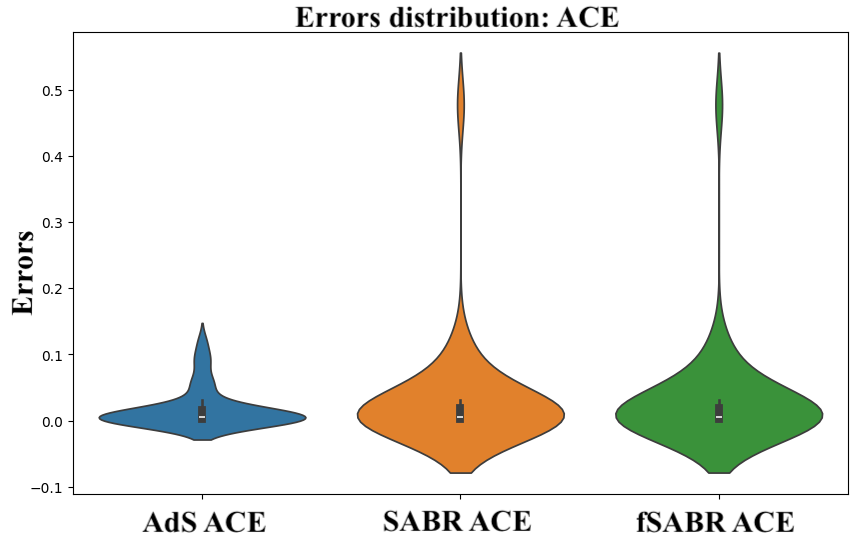}
    \caption{\small Violin plot ACE}
    \label{fig:ACE_errors}
  \end{subfigure}
  \caption{\small Violin plots for MSE, MAE, RMSCE, and ACE error metrics.}
  \label{fig:error_metrics}
\end{figure}

The SABR and fSABR models, as highlighted in the literature, exhibit higher errors for OTM options, as confirmed by error metrics and fitting graphs. ITM and OTM options are sensitive to complex market dynamics, with their implied volatility reflecting extreme conditions and asset price expectations. The AdS model demonstrates superior performance due to its dynamic flexibility, allowing the Hurst exponent to adjust based on moneyness. In ITM and OTM regions, H decreases, creating an inverse smile effect that captures deviations from the EMH and models structural inefficiencies and unique implied volatility patterns observed far from the ATM region.

The AdS model aligns with the EMH ($H \sim 1/2$) in the ATM region, consistent with traditional models like SABR, but it shows $H < 1/2$ in ITM and OTM regions. This indicates weaker memory and more stochastic price behavior, effectively capturing implied volatility variations. The $H(\frac{S}{K})$ function in the AdS model dynamically adjusts to moneyness, enabling it to account for market inefficiencies and shifts in structure and investor behavior more effectively than the static SABR and fSABR models.

\section{Conclusion} \label{conclusion}
This research presents an implied volatility model that incorporates long-term memory, represented by the Hurst exponent $H$, and sensitivity to moneyness. Using a closed-form formula, the model outperformed the SABR and fSABR models in fitting market data. It expresses implied volatility as a function of the distance between the asset price and the strike price, capturing the concavity of the volatility surface and long-term memory effects.

The model introduces a nonlinear relationship between volatility and moneyness, with a quadratic term to reflect typical volatility surface behaviors, particularly in ITM and OTM regions. The Hurst exponent $H$ varies with moneyness, adding a memory effect that influences volatility dynamics. Calibration using the Optuna framework minimized fitting errors, showing superior performance in error metrics (RMSE, MAE, MSE) and curvature metrics (ACE, RMSCE), particularly in extreme moneyness regions with high surface concavity.

A key feature is its ability to capture implied regularity around the ATM region ($\frac{S}{K}=1$), where $H=1/2$, consistent with maximum temporal uncertainty and Brownian motion behavior. The model also satisfies the no-arbitrage condition, ensuring that predicted volatility surfaces align with financial theory and market efficiency by preventing risk-free profit opportunities.
\newline

\textbf{Declaration of generative AI and AI-assisted technologies in the writing process.}
During the preparation of this work the authors used ChatGPT in order to improve the readability and language of the manuscript.

\section{Annex} \label{sec:Annex}
Let the option price of a call 
\begin{equation}
    C\left(t,K;\sigma\right) = S_t\mathcal{N}\left(d_1\left(S_t,K,r,T,t,\sigma(K)\right)\right) - Ke^{-r(T-t)}\mathcal{N}\left(d_2\left(S_t,K,r,T,t,\sigma(K)\right)\right),
\end{equation}
where $S_t$ is the underlying price, $K$ the strike price, $r$ the free-rate risk neutral, $T$ the expiration date, $t$ the current date and $\sigma(K)$ the implied volatility respectively. In particular $\sigma(K)$ is described by the AdS model
\begin{equation}
    \sigma(K) = \alpha\left(\frac{S_t}{K}-\frac{S_t}{K_{\min}}\right)^2 e^{-\beta H(K) \left(\frac{S_t}{K}-\frac{S_t}{K_{\min}}\right)} + \epsilon,
\end{equation}
with the Hurst exponent
\begin{equation}
    H(K) = \frac{1}{2} \left( 1 + \left| 1 - \frac{S}{K_{\min}} \right|^\delta \right) \frac{1}{\left( 1 + \left| \frac{S}{K} - \frac{S}{K_{\min}} \right|^\delta \right)}
\end{equation}
with $\alpha, \epsilon >0$.
We need to introduce some conditions on our model to make financial sense of it. In this way we will obtain defining intervals for the parameters $\beta$ and $\delta$. In the second analysis we will then have to place conditions on the call price.
\subsection{H's first derivative}
In order to have concavity we need to impose that
\begin{equation}\label{eq:condition H}
    \begin{cases}
        \frac{dH\left(1/K\right)}{d(1/K)} > 0 \text{  for  }\frac{1}{K}<\frac{1}{K_{\min}},\\
        \frac{dH\left(1/K\right)}{d(1/K)} < 0 \text{  for  }\frac{1}{K}>\frac{1}{K_{\min}}.
    \end{cases}
\end{equation}
Defining $g\left(\frac{1}{K}\right) = \frac{1}{K}-\frac{1}{K_{\min}}$ we have
\begin{equation}\label{eq:dH}
    \frac{dH(1/K)}{d(1/K)} = -\text{sgn}(g)\frac{H(1/K)\delta S_t^\delta\vert g\vert^{\delta-1}}{1+S_t^\delta\vert g\vert^\delta}.
\end{equation}
The equation \eqref{eq:dH} satisfies the conditions in equation \eqref{eq:condition H} only for $\delta > 0$.
\subsection{\texorpdfstring{$\sigma$}{Lg}'s first derivative}
In order to have convexity we need to impose
\begin{equation}\label{eq: condition sigma}
    \begin{cases}
        \frac{d\sigma\left(1/K\right)}{d(1/K)} > 0 \text{  for  }\frac{1}{K}>\frac{1}{K_{\min}},\\
        \frac{d\sigma\left(1/K\right)}{d(1/K)} < 0 \text{  for  }\frac{1}{K}<\frac{1}{K_{\min}}.
    \end{cases}
\end{equation}
The first derivative of the implied volatility is:
\begin{equation}
    \frac{d\sigma(1/K)}{d(1/K)} = \sigma\left\{\frac{2}{g\left(\frac{1}{K}\right)} - H\left(\frac{1}{K}\right)\beta S_t\left[\frac{1+S_t^\delta \vert g\vert^\delta(1-\delta)}{1 + S_t^\delta\vert g\vert^\delta}\right]\right\}.
\end{equation}
Using the conditions on equation \eqref{eq: condition sigma} we can obtain the following interval for $\beta$:
\begin{equation}
    \vert\beta\vert < \frac{2}{H(1/K)S_t\vert g\vert}\frac{1+S_t^\delta \vert g\vert^\delta}{1+S_t^\delta\vert g\vert^{\delta-1}(1-\delta)}.
\end{equation}
For $\beta$ values to be admissible, we must require the second member to be positive. In particular, that the term $\frac{1+S_t^\delta\vert g\vert^\delta}{1 + S_t^\delta\vert g\vert^{\delta-1}(1-\delta)}$ is. We need to impose that
\begin{equation}\label{eq: delta condition}
    (\delta - 1)  < \frac{1}{\left(S_t\vert g\vert\right)^\delta}.
\end{equation}
Because of the arbitrariness of the value of $S_t\vert g\vert$, we choose the $\delta$ parameter as conservatively as possible, that is, when $S_t\vert g\vert\to\infty$. Therefore, condition \eqref{eq: delta condition} is satisfied for $0<\delta<1$.
\subsection{Arbitrage-free volatility surface}\label{theorem: arbitrage}
Given a set of parameters $\overline{p} = \{\alpha,\beta,\delta,\epsilon\}$, let $\sigma(T,K;\overline{p})$ be a parametrized implied volatility surface defined on $\Pi = \Pi_T \times \Pi_K = \left(t_0,\infty\right)\times\mathbb{R}_+$. Let $C(T,K;\overline{p})$ be its call pricing function, which is extended to the limit points at $T=t_0$. The parametrization is called free of “butterfly” arbitrage if the following conditions hold on the call pricing function and a constant $s > 0$:
\begin{itemize}
    \item \textit{i}) $C(T,\cdot;\overline{p})$ is convex and non-increasing for all $T\in \Pi_T$;
    \item \textit{ii}) $\lim\limits_{K\to\infty} C(T,K;\overline{p})=0$ for all $T\in \Pi_T$;
    \item \textit{iii}) $\left(s-K\right)^+\leq C(T,K;\overline{p})\leq s$ for all $(T,K)\in\Pi$;
    \item \textit{iv}) $C(t_0,K;\overline{p})=(s-K)^+$ for all $K\in\Pi_K$.
\end{itemize}
    If the following additional condition holds, the pricing surface is also free of “calendar” arbitrage, and we call the surface arbitrage-free:
\begin{itemize}
    \item \textit{v}) $C(\cdot,K;\overline{p})$ is non-decreasing for all $K\in\Pi_K$.
\end{itemize}
\textit{Proof}\\
\begin{itemize}
    \item \textit{i}) For any $T\in\Pi_T$, $\frac{\partial C}{\partial K}\leq 0$. In fact 
    \[\frac{\partial C}{\partial K} = S_t\phi(d_1)\frac{\partial d_1}{\partial K}-e^{-r(T-t)}\mathcal{N}(d_2)-Ke^{-r(T-t)}\phi(d_2)\frac{\partial d_2}{\partial K}\]
    \[=-e^{-r(T-t)}\mathcal{N}(d_2)+\left[S_t\phi(d_1)-Ke^{-r(T-t)}\phi(d_2)\right]\frac{\partial d_1}{\partial K}+K\sqrt{T-t}e^{-r(T-t)}\phi(d_2)\frac{\partial \sigma}{\partial K}.\]
    From $\left[S_t\phi(d_1)-Ke^{-r(T-t)}\phi(d_2)\right]=0$ we have
    \[\frac{\partial C}{\partial K} = -e^{-r(T-t)}\mathcal{N}(d_2)-\frac{1}{K}\sqrt{T-t}e^{-r(T-t)}\phi(d_2)\frac{d \sigma}{d 1/K}\]
The first term is always non-positive. For the second derivative term we need to split the problem in two cases. When $K<K_{\min}$ the $\frac{d\sigma}{d1/K}>0$ we have no problem and the first derivative of the price $C$ is always non-positive. When $K>K_{\min}$ the $\frac{d\sigma}{d1/K}<0$, therefore we need:
\[0>\frac{d\sigma}{d1/K} \geq -\frac{K\mathcal{N}(d_2)}{\phi(d_2)\sqrt{T-t}}.\]
       About the convexity of price $C$ we need to require $\frac{\partial^2 C}{\partial K^2}>0$. The condition is
       \[-\phi(d_2)\left[\frac{\partial d_1}{\partial K}+K\frac{d^2\sigma}{d K^2}\right]+K\sqrt{T-t}\frac{d\sigma}{dK}\frac{\partial \phi(d_2)}{\partial K}>0.
       \]
\noindent
Figure \ref{fig:dC_dK} shows the computational study conducted on the behavior on the first and second derivatives conditions, with respect to the parameter ranges $\beta \in[-1,1]$, $\alpha \in[0,1]$, and $\delta \in[0,1]$.

\begin{figure}[!ht]
    \centering

    \begin{subfigure}[b]{0.7\textwidth}
        \centering
        \includegraphics[width=\textwidth]{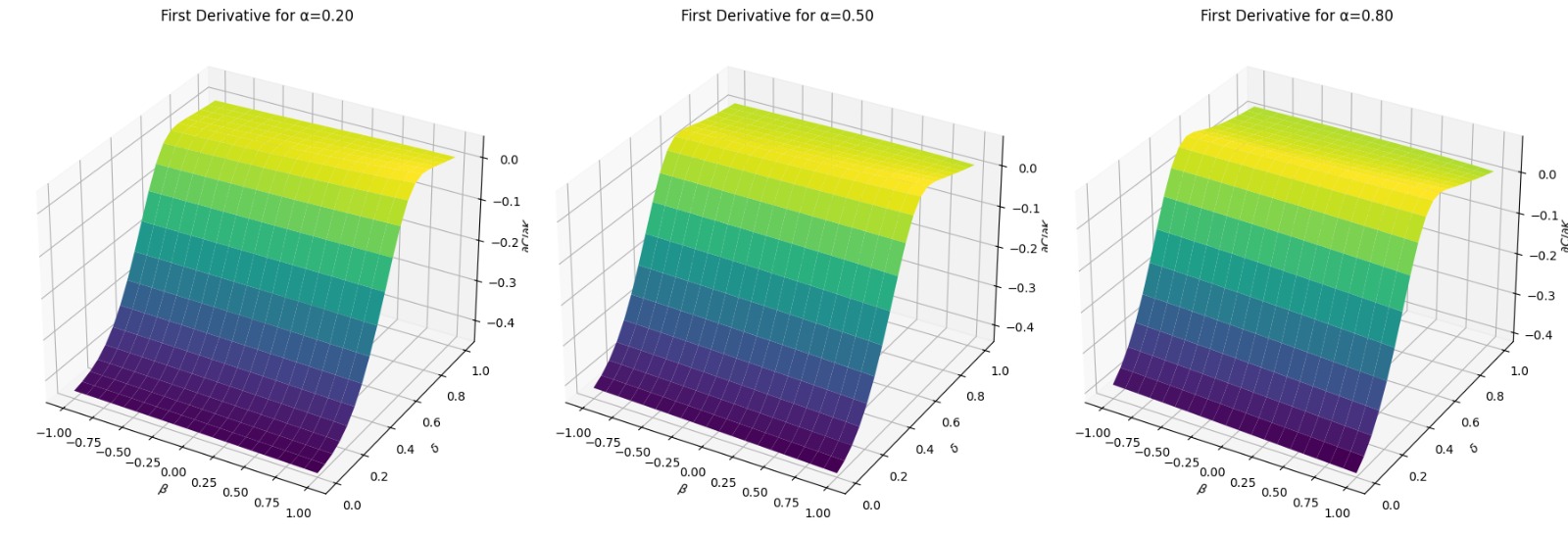}
        \caption{$\frac{\partial C}{\partial K}$}
        \label{fig:grad1}
    \end{subfigure}
    \hfill
    \begin{subfigure}[b]{0.7\textwidth}
        \centering
        \includegraphics[width=\textwidth]{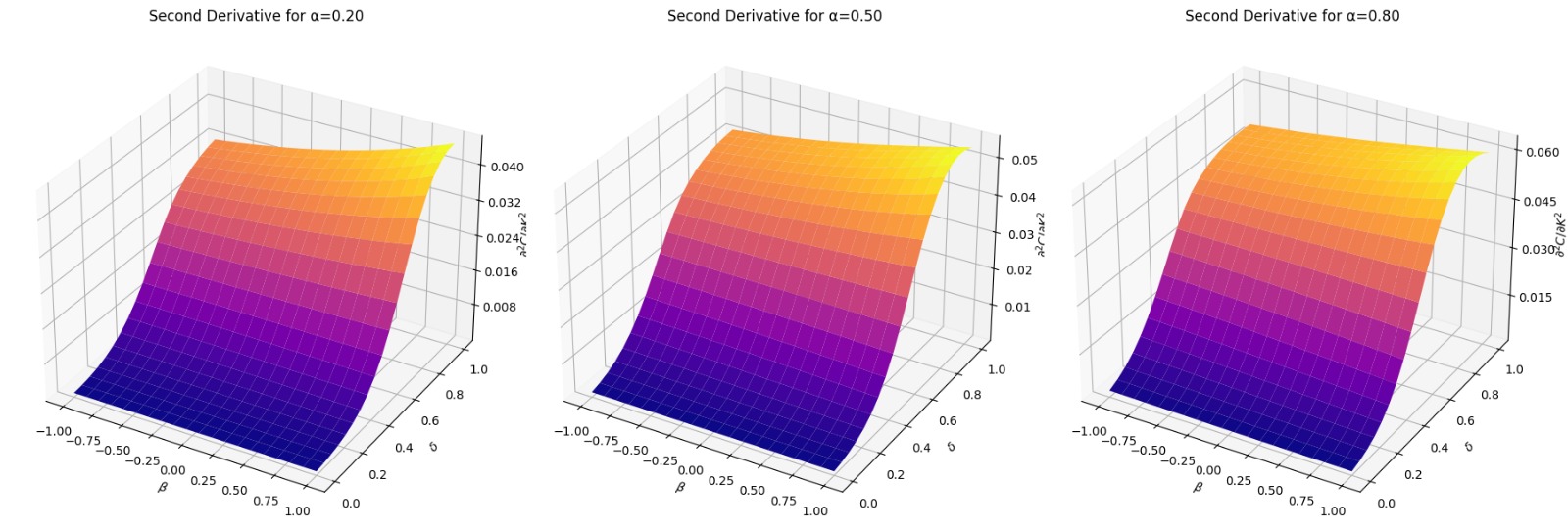}
        \caption{$\frac{\partial^2 C}{\partial K^2}$}
        \label{fig:grad2}
    \end{subfigure}

    \caption{Parameters condition}
    \label{fig:dC_dK}
\end{figure}

        \item \textit{ii}) For all $T\in\Pi_T$
        \[\lim\limits_{K\to+\infty}C(T,K;\overline{p}) = \frac{S_t}{\sqrt{2\pi}}\int\limits_{-\infty}^{\lim\limits_{K\to+\infty}d_1}e^{-\frac{x^2}{2}}dx -e^{-r(T-t)}\lim\limits_{K\to+\infty}\frac{K}{\sqrt{2\pi}}\int\limits_{-\infty}^{d_1-\sigma\sqrt{T-t}}e^{-\frac{x^2}{2}}dx.\]
        To compute $\lim\limits_{k\to+\infty}d_1$ we need $\lim\limits_{k\to+\infty}\sigma(K)$ and $\lim\limits_{k\to+\infty}H(K)$:
        \[H_\infty = \lim\limits_{k\to+\infty}H(K)=\frac{1}{2}\left(1+\left\vert1-\frac{S_t}{K_{\min}}\right\vert^\delta\right)\frac{1}{1+\left(\frac{S_t}{K_{\min}}\right)^\delta},\]
        \[\sigma_\infty = \lim\limits_{k\to+\infty}\sigma(K)=\alpha\frac{S_t^2}{K_{\min}^2}e^{\beta\frac{S_t}{K_{\min}}H_\infty}+\epsilon\]
        Therefore
        \[\lim\limits_{K\to+\infty}d_1=\frac{\lim\limits_{K\to+\infty}\ln\left(\frac{S_t}{K}\right)+\left(r+\frac{1}{2}\sigma_{\infty}^2\right)(T-t)}{\sigma_\infty\sqrt{T-t}}=-\infty.\]
        Finally we get $\lim\limits_{k\to+\infty}C(T,K;\overline{p})=0$.
        \item \textit{iii}) We want to proof that for all $T,K\in\Pi$ and for any $s>0$ constant we can write 
        \[C(T,K;\overline{p}) = s\mathcal{N}(d_1)-Ke^{-r(T-t)}\mathcal{N}(d_2)\leq s.\]
        From this inequality we obtain a trivial relation
        \[\mathcal{N}(d_1)-\frac{K}{s}e^{-r(T-t)}\mathcal{N}(d_2)\leq 1.\]
        In fact $1\geq\mathcal{N}(d1)\geq\mathcal{N}(d_2)\geq0$ and $K,s,r>0$, $T\geq t$.\\
        To proof $(s-K)^+\leq C(T,K;\overline{p})$, we can write the European call price at time $T$ as the discounted expectation under the risk-neutral measure:
\begin{equation}
C(T, K; \overline{p}) = e^{-r(T - t)} \mathbb{E}^\mathbb{Q}[(S_T - K)^+],
\end{equation}
where:
\begin{itemize}
    \item \( S_T \) is the price of the underlying asset at time \( T \),
    \item \( r \) is the risk-free interest rate,
    \item \( \mathbb{Q} \) is the risk-neutral measure,
    \item \( (S_T - K)^+ \) is the payoff of the call option at maturity \( T \).
\end{itemize}
By definition, $(S_T-K)^+\geq(s-K)^+$, because the underlying price $S_T$ can only stay the same or increase over time. 
Taking the expectation of this inequality, we have
\[C(T,K;\overline{p})=\mathbb{E}\left[ (S_T-K)^+\right]\geq (s-K)^+.\]

        \item \textit{iv}) The price of \( C(T, K; \overline{p}) \) at time \( T \) is given by the discounted expectation under the risk-neutral measure:
\[
C(T, K; \overline{p}) = e^{-r(T - t_0)} \mathbb{E}^\mathbb{Q}[(S_T - K)^+],
\]

At the initial time \( t_0 \), the price is known and equals \( S_{t_0} = s \). Therefore, there is no uncertainty about the payoff, and the formula simplifies to:
\[
C(t_0, K; \overline{p}) = \mathbb{E}^\mathbb{Q}[(s - K)^+] = (s - K)^+.
\]
This equality shows that the call price at the initial time is determined exclusively by the immediate payoff.

The underlying price process \( S_t \) is assumed to follow a multifractional Brownian motion (mBm), where the Hurst parameter \( H(t) \) satisfies:
\[
\lim_{t \to T} H(t) = \frac{1}{2}.
\]
This condition ensures that $S_t$ is a martingale, making it consistent with the no-arbitrage condition and risk-neutral pricing framework. Specifically, as $H(t) \to \frac{1}{2}$, the multifractional Brownian motion (mBm) aligns with geometric Brownian motion, preserving the properties required for a martingale under the risk-neutral measure $\mathbb{Q}$. For more information see ~\cite{disciorio2020option}.
        \item \textit{v}) For all $K\in\Pi_K$ we need $\frac{\partial C(T,K;\overline{p})}{\partial T}\geq0$.
        \[\frac{\partial C}{\partial T} = S_t\frac{\partial\mathcal{N}(d_1)}{\partial T} + rKe^{-r(T-t)}\mathcal{N}(d_2)-Ke^{-r(T-t)}\frac{\partial\mathcal{N}(d_2)}{\partial T}\]
        \[=S_t\phi(d_1)\frac{\partial d_1}{\partial T}+rKe^{-r(T-t)}\mathcal{N}(d_2)-Ke^{-r(T-t)}\phi(d_2)\frac{\partial d_1}{\partial T}+\frac{\sigma Ke^{-r(T-t)}}{2\sqrt{T-t}}\phi(d_2)\]
        \[=rKe^{-r(T-t)}\mathcal{N}(d_2)+\frac{\sigma Ke^{-r(T-t)}}{2\sqrt{T-t}}\phi(d_2)+\left[S_t\phi(d_1)-Ke^{-r(T-t)}\phi(d_2)\right]\frac{\partial d_1}{\partial T}.\]
        Since $\left[S_t\phi(d_1)-Ke^{-r(T-t)}\phi(d_2)\right]$ cancels and the terms $rKe^{-r(T-t)}\mathcal{N}(d_2)$ and $\frac{\sigma Ke^{-r(T-t)}}{2\sqrt{T-t}}\phi(d_2)$ are non-negative, we conclude that $\frac{\partial C}{\partial T}\geq0$.
\end{itemize}
\newgeometry{left=0.125cm, right=0.25cm, top=0.5cm, bottom=0.5cm}
\subsection{Table}\label{sec:data}
\begin{table}[H] 
    \centering
    \fontsize{5}{10}\selectfont
    \begin{tabular}{|l|l|l|l|l|l|l|l|l|l|l|l|l|} \hline  
        & \multicolumn{4}{|c|}{AdS model} & \multicolumn{4}{|c|}{SABR} & \multicolumn{4}{|c|}{fSABR} \\ \hline  
        Ticker & MSE & MAE & RMSCE & ACE & MSE & MAE & RMSCE & ACE & MSE & MAE & RMSCE & ACE \\ \hline  
        AMZN  & $4.00\cdot10^{-4}$ & $1.49\cdot10^{-2}$ & $3.42\cdot10^{-3}$ & $1.94\cdot10^{-3}$ & $6.00\cdot10^{-4}$ & $2.01\cdot10^{-2}$ & $3.44\cdot10^{-3}$ & $1.92\cdot10^{-3}$ & $4.62\cdot10^{-2}$ & $2.05\cdot10^{-1}$ & $3.43\cdot10^{-3}$ & $1.89\cdot10^{-3}$ \\  
        TSLA  & $4.66\cdot10^{-2}$ & $1.48\cdot10^{-1}$ & $1.30\cdot10^{-3}$ & $5.00\cdot10^{-4}$ & $9.54\cdot10^{-2}$ & $2.19\cdot10^{-1}$ & $1.30\cdot10^{-3}$ & $5.00\cdot10^{-4}$ & $1.20\cdot10^{0}$ & $8.51\cdot10^{-1}$ & $1.30\cdot10^{-3}$ & $5.00\cdot10^{-4}$ \\  
        STLA  & $1.47\cdot10^{-1}$ & $1.72\cdot10^{-1}$ & $1.22\cdot10^{0}$ & $6.27\cdot10^{-1}$ & $1.59\cdot10^{-1}$ & $2.22\cdot10^{-1}$ & $1.26\cdot10^{0}$ & $6.43\cdot10^{-1}$ & $4.99\cdot10^{-1}$ & $6.02\cdot10^{-1}$ & $1.26\cdot10^{0}$ & $6.49\cdot10^{-1}$ \\  
        MS    & $3.53\cdot10^{-2}$ & $1.12\cdot10^{-1}$ & $5.42\cdot10^{-2}$ & $2.43\cdot10^{-2}$ & $4.71\cdot10^{-2}$ & $1.62\cdot10^{-1}$ & $5.46\cdot10^{-2}$ & $2.56\cdot10^{-2}$ & $7.63\cdot10^{-1}$ & $6.17\cdot10^{-1}$ & $5.35\cdot10^{-2}$ & $2.42\cdot10^{-2}$ \\  
        NKE   & $1.07\cdot10^{-2}$ & $4.08\cdot10^{-2}$ & $2.72\cdot10^{-2}$ & $1.01\cdot10^{-2}$ & $3.37\cdot10^{-2}$ & $1.40\cdot10^{-1}$ & $2.72\cdot10^{-2}$ & $1.04\cdot10^{-2}$ & $4.08\cdot10^{-1}$ & $5.65\cdot10^{-1}$ & $2.61\cdot10^{-2}$ & $9.93\cdot10^{-3}$ \\  
        META  & $6.31\cdot10^{-3}$ & $5.82\cdot10^{-2}$ & $1.72\cdot10^{-3}$ & $7.19\cdot10^{-4}$ & $8.28\cdot10^{-3}$ & $6.85\cdot10^{-2}$ & $1.72\cdot10^{-3}$ & $7.20\cdot10^{-4}$ & $5.43\cdot10^{-2}$ & $1.74\cdot10^{-1}$ & $1.72\cdot10^{-3}$ & $7.18\cdot10^{-4}$ \\  
        GOOGL & $9.95\cdot10^{-2}$ & $1.89\cdot10^{-1}$ & $3.48\cdot10^{-2}$ & $1.55\cdot10^{-2}$ & $7.68\cdot10^{-2}$ & $2.35\cdot10^{-1}$ & $3.71\cdot10^{-2}$ & $1.66\cdot10^{-2}$ & $1.59\cdot10^{0}$ & $6.11\cdot10^{-1}$ & $1.34\cdot10^{-1}$ & $3.35\cdot10^{-2}$ \\  
        NVDA  & $6.59\cdot10^{0}$ & $1.57\cdot10^{0}$ & $2.50\cdot10^{-2}$ & $5.90\cdot10^{-3}$ & $9.44\cdot10^{0}$ & $1.91\cdot10^{0}$ & $2.39\cdot10^{-2}$ & $5.60\cdot10^{-3}$ & $5.70\cdot10^{1}$ & $5.13\cdot10^{0}$ & $2.69\cdot10^{-2}$ & $5.40\cdot10^{-3}$ \\  
        MSFT  & $1.63\cdot10^{-2}$ & $9.25\cdot10^{-2}$ & $1.05\cdot10^{-2}$ & $5.88\cdot10^{-3}$ & $2.06\cdot10^{-2}$ & $1.28\cdot10^{-1}$ & $1.06\cdot10^{-2}$ & $6.02\cdot10^{-3}$ & $1.66\cdot10^{-1}$ & $2.83\cdot10^{-1}$ & $1.39\cdot10^{-2}$ & $7.07\cdot10^{-3}$ \\  
        ACN   & $5.00\cdot10^{-4}$ & $1.56\cdot10^{-2}$ & $1.11\cdot10^{-2}$ & $4.20\cdot10^{-3}$ & $1.40\cdot10^{-3}$ & $2.50\cdot10^{-2}$ & $8.50\cdot10^{-3}$ & $3.70\cdot10^{-3}$ & $1.46\cdot10^{-1}$ & $3.75\cdot10^{-1}$ & $1.84\cdot10^{-2}$ & $5.70\cdot10^{-3}$ \\  
        NFLX  & $2.81\cdot10^{-3}$ & $4.25\cdot10^{-2}$ & $5.40\cdot10^{-4}$ & $2.37\cdot10^{-4}$ & $3.27\cdot10^{-3}$ & $4.04\cdot10^{-2}$ & $5.51\cdot10^{-4}$ & $2.37\cdot10^{-4}$ & $2.74\cdot10^{-2}$ & $1.24\cdot10^{-1}$ & $5.60\cdot10^{-4}$ & $2.36\cdot10^{-4}$ \\  
        MRVL  & $2.34\cdot10^{-3}$ & $3.82\cdot10^{-2}$ & $3.48\cdot10^{-2}$ & $1.77\cdot10^{-2}$ & $3.15\cdot10^{-2}$ & $1.42\cdot10^{-1}$ & $1.27\cdot10^{-1}$ & $4.01\cdot10^{-2}$ & $2.51\cdot10^{-1}$ & $1.99\cdot10^{-1}$ & $4.95\cdot10^{-1}$ & $1.15\cdot10^{-1}$ \\  
        SPY   & $1.83\cdot10^{-2}$ & $1.17\cdot10^{-1}$ & $3.59\cdot10^{-3}$ & $1.27\cdot10^{-3}$ & $3.30\cdot10^{-2}$ & $1.57\cdot10^{-1}$ & $5.44\cdot10^{-3}$ & $2.10\cdot10^{-3}$ & $5.10\cdot10^{-1}$ & $6.07\cdot10^{-1}$ & $5.63\cdot10^{-3}$ & $2.22\cdot10^{-3}$ \\  
  RUT   & $6.80\cdot10^{-3}$ & $6.06\cdot10^{-2}$ & $1.40\cdot10^{-3}$ & $4.00\cdot10^{-4}$ & $1.40\cdot10^{-2}$ & $9.60\cdot10^{-2}$ & $1.00\cdot10^{-3}$ & $4.00\cdot10^{-4}$ & $2.06\cdot10^{-1}$ & $3.63\cdot10^{-1}$ & $1.40\cdot10^{-3}$ & $5.00\cdot10^{-4}$ \\  
        SPX   & $8.48\cdot10^{-2}$ & $2.14\cdot10^{-1}$ & $7.31\cdot10^{-4}$ & $3.37\cdot10^{-4}$ & $1.05\cdot10^{-1}$ & $1.42\cdot10^{-1}$ & $7.35\cdot10^{-4}$ & $3.39\cdot10^{-4}$ & $5.47\cdot10^{-1}$ & $5.82\cdot10^{3}$ & $7.29\cdot10^{-4}$ & $3.37\cdot10^{-4}$ \\  
        KO    & $3.64\cdot10^{-2}$ & $1.35\cdot10^{-1}$ & $2.41\cdot10^{-1}$ & $8.76\cdot10^{-2}$ & $6.90\cdot10^{-2}$ & $1.83\cdot10^{-1}$ & $9.03\cdot10^{-2}$ & $4.89\cdot10^{-2}$ & $4.48\cdot10^{-1}$ & $4.45\cdot10^{-1}$ & $3.01\cdot10^{-1}$ & $1.02\cdot10^{-1}$ \\  
        BAC   & $1.10\cdot10^{-3}$ & $2.61\cdot10^{-2}$ & $9.59\cdot10^{-2}$ & $6.27\cdot10^{-2}$ & $1.70\cdot10^{-3}$ & $3.03\cdot10^{-2}$ & $9.63\cdot10^{-2}$ & $6.20\cdot10^{-2}$ & $9.80\cdot10^{-3}$ & $8.30\cdot10^{-2}$ & $9.60\cdot10^{-2}$ & $6.13\cdot10^{-2}$ \\  
        GD    & $5.00\cdot10^{-4}$ & $1.71\cdot10^{-2}$ & $4.00\cdot10^{-3}$ & $2.30\cdot10^{-3}$ & $6.00\cdot10^{-4}$ & $1.97\cdot10^{-2}$ & $4.00\cdot10^{-3}$ & $2.20\cdot10^{-3}$ & $4.10\cdot10^{-3}$ & $5.87\cdot10^{-2}$ & $4.00\cdot10^{-3}$ & $2.20\cdot10^{-3}$ \\  
        AAPL  & $8.00\cdot10^{-4}$ & $2.16\cdot10^{-2}$ & $5.48\cdot10^{-2}$ & $1.19\cdot10^{-2}$ & $9.00\cdot10^{-4}$ & $2.43\cdot10^{-2}$ & $3.50\cdot10^{-3}$ & $1.90\cdot10^{-3}$ & $4.20\cdot10^{-3}$ & $5.91\cdot10^{-2}$ & $5.16\cdot10^{-2}$ & $1.12\cdot10^{-2}$ \\  
        GM    & $4.73\cdot10^{-3}$ & $5.82\cdot10^{-2}$ & $1.21\cdot10^{-1}$ & $5.52\cdot10^{-2}$ & $2.76\cdot10^{-1}$ & $3.58\cdot10^{-1}$ & $6.84\cdot10^{-1}$ & $2.27\cdot10^{-1}$ & $1.56\cdot10^{0}$ & $3.74\cdot10^{-1}$ & $1.31\cdot10^{0}$ & $3.43\cdot10^{-1}$ \\  
        GS    & $1.46\cdot10^{-3}$ & $3.11\cdot10^{-2}$ & $2.83\cdot10^{-3}$ & $1.10\cdot10^{-3}$ & $3.26\cdot10^{-2}$ & $1.30\cdot10^{-1}$ & $2.48\cdot10^{-2}$ & $5.50\cdot10^{-3}$ & $3.48\cdot10^{-1}$ & $1.86\cdot10^{-1}$ & $9.34\cdot10^{-2}$ & $1.82\cdot10^{-2}$ \\  
        IBM   & $7.70\cdot10^{-3}$ & $6.33\cdot10^{-2}$ & $8.74\cdot10^{-3}$ & $4.89\cdot10^{-3}$ & $1.75\cdot10^{-2}$ & $1.21\cdot10^{-1}$ & $8.74\cdot10^{-3}$ & $4.90\cdot10^{-3}$ & $1.25\cdot10^{-1}$ & $2.34\cdot10^{-1}$ & $8.79\cdot10^{-3}$ & $5.02\cdot10^{-3}$ \\  
        INTC  & $6.20\cdot10^{-3}$ & $6.41\cdot10^{-2}$ & $2.58\cdot10^{-1}$ & $9.61\cdot10^{-2}$ & $9.60\cdot10^{-3}$ & $7.65\cdot10^{-2}$ & $2.58\cdot10^{-1}$ & $9.26\cdot10^{-2}$ & $2.40\cdot10^{-2}$ & $1.32\cdot10^{-1}$ & $2.58\cdot10^{-1}$ & $9.25\cdot10^{-2}$ \\  
        JPM   & $3.30\cdot10^{-3}$ & $4.58\cdot10^{-2}$ & $4.00\cdot10^{-3}$ & $2.30\cdot10^{-3}$ & $6.60\cdot10^{-3}$ & $5.96\cdot10^{-2}$ & $3.90\cdot10^{-3}$ & $2.10\cdot10^{-3}$ & $1.04\cdot10^{-1}$ & $1.39\cdot10^{-1}$ & $3.90\cdot10^{-3}$ & $2.10\cdot10^{-3}$ \\  
        MA    & $3.09\cdot10^{-2}$ & $1.17\cdot10^{-1}$ & $1.99\cdot10^{-2}$ & $4.80\cdot10^{-3}$ & $1.13\cdot10^{-2}$ & $9.07\cdot10^{-2}$ & $8.70\cdot10^{-3}$ & $2.90\cdot10^{-3}$ & $1.22\cdot10^{-1}$ & $2.12\cdot10^{-1}$ & $2.82\cdot10^{-2}$ & $6.40\cdot10^{-3}$ \\  
        MCD   & $3.58\cdot10^{-2}$ & $1.18\cdot10^{-1}$ & $3.09\cdot10^{-2}$ & $8.90\cdot10^{-3}$ & $1.09\cdot10^{-2}$ & $7.70\cdot10^{-2}$ & $4.30\cdot10^{-3}$ & $2.40\cdot10^{-3}$ & $3.01\cdot10^{-1}$ & $2.37\cdot10^{-1}$ & $3.72\cdot10^{-2}$ & $1.03\cdot10^{-2}$ \\  
 MRNA  & $3.60\cdot10^{-3}$ & $4.93\cdot10^{-2}$ & $3.03\cdot10^{-3}$ & $1.24\cdot10^{-3}$ & $7.10\cdot10^{-3}$ & $6.70\cdot10^{-2}$ & $3.05\cdot10^{-3}$ & $1.38\cdot10^{-3}$ & $2.14\cdot10^{-2}$ & $1.28\cdot10^{-1}$ & $3.05\cdot10^{-3}$ & $1.27\cdot10^{-3}$ \\  
        MSCI  & $4.49\cdot10^{-2}$ & $1.51\cdot10^{-1}$ & $3.03\cdot10^{-3}$ & $1.24\cdot10^{-3}$ & $6.25\cdot10^{-2}$ & $2.01\cdot10^{-1}$ & $3.08\cdot10^{-3}$ & $1.42\cdot10^{-3}$ & $2.01\cdot10^{-1}$ & $3.12\cdot10^{-1}$ & $3.04\cdot10^{-3}$ & $1.29\cdot10^{-3}$ \\  
        NDAQ  & $8.14\cdot10^{-4}$ & $2.45\cdot10^{-2}$ & $8.17\cdot10^{-3}$ & $5.78\cdot10^{-3}$ & $1.92\cdot10^{-2}$ & $1.23\cdot10^{-1}$ & $1.18\cdot10^{-2}$ & $7.87\cdot10^{-3}$ & $1.14\cdot10^{0}$ & $5.11\cdot10^{-1}$ & $1.78\cdot10^{-1}$ & $8.26\cdot10^{-2}$ \\  
        PFE   & $3.60\cdot10^{-3}$ & $4.94\cdot10^{-2}$ & $2.04\cdot10^{-1}$ & $1.19\cdot10^{-1}$ & $3.90\cdot10^{-3}$ & $5.15\cdot10^{-2}$ & $2.03\cdot10^{-1}$ & $1.17\cdot10^{-1}$ & $2.46\cdot10^{-2}$ & $1.45\cdot10^{-1}$ & $2.03\cdot10^{-1}$ & $1.16\cdot10^{-1}$ \\  
        SBUX  & $2.00\cdot10^{-3}$ & $3.73\cdot10^{-2}$ & $2.27\cdot10^{-2}$ & $1.21\cdot10^{-2}$ & $7.20\cdot10^{-3}$ & $5.42\cdot10^{-2}$ & $1.17\cdot10^{-1}$ & $3.91\cdot10^{-2}$ & $1.48\cdot10^{-1}$ & $1.67\cdot10^{-1}$ & $5.26\cdot10^{-1}$ & $1.29\cdot10^{-1}$ \\  
        UBER  & $1.08\cdot10^{-2}$ & $6.13\cdot10^{-2}$ & $9.13\cdot10^{-2}$ & $3.16\cdot10^{-2}$ & $1.39\cdot10^{-2}$ & $6.13\cdot10^{-2}$ & $9.16\cdot10^{-2}$ & $3.34\cdot10^{-2}$ & $4.57\cdot10^{-2}$ & $1.23\cdot10^{-1}$ & $9.44\cdot10^{-2}$ & $3.02\cdot10^{-2}$ \\  
        V     & $1.60\cdot10^{-2}$ & $8.16\cdot10^{-2}$ & $8.13\cdot10^{-3}$ & $3.76\cdot10^{-3}$ & $3.21\cdot10^{-2}$ & $1.27\cdot10^{-1}$ & $8.40\cdot10^{-3}$ & $3.97\cdot10^{-3}$ & $5.79\cdot10^{-1}$ & $4.27\cdot10^{-1}$ & $1.95\cdot10^{-2}$ & $6.03\cdot10^{-3}$ \\  
        AVGO  & $1.27\cdot10^{1}$ & $2.04\cdot10^{0}$ & $1.00\cdot10^{0}$ & $1.47\cdot10^{-1}$ & $1.63\cdot10^{1}$ & $2.29\cdot10^{0}$ & $4.49\cdot10^{-1}$ & $8.41\cdot10^{-2}$ & $6.85\cdot10^{1}$ & $5.10\cdot10^{0}$ & $1.28\cdot10^{0}$ & $1.84\cdot10^{-1}$ \\  
        JNJ   & $8.80\cdot10^{-3}$ & $5.36\cdot10^{-2}$ & $1.32\cdot10^{-2}$ & $8.00\cdot10^{-3}$ & $1.10\cdot10^{-2}$ & $5.37\cdot10^{-2}$ & $1.32\cdot10^{-2}$ & $7.90\cdot10^{-3}$ & $2.46\cdot10^{-2}$ & $8.09\cdot10^{-2}$ & $1.32\cdot10^{-2}$ & $7.90\cdot10^{-3}$ \\  
        XOM   & $1.10\cdot10^{-1}$ & $1.66\cdot10^{-1}$ & $2.45\cdot10^{-1}$ & $5.05\cdot10^{-2}$ & $1.51\cdot10^{-1}$ & $1.98\cdot10^{-1}$ & $3.81\cdot10^{-1}$ & $7.40\cdot10^{-2}$ & $2.37\cdot10^{-1}$ & $2.44\cdot10^{-1}$ & $4.83\cdot10^{-1}$ & $9.07\cdot10^{-2}$ \\  
        UNH   & $7.20\cdot10^{-3}$ & $5.02\cdot10^{-2}$ & $8.69\cdot10^{-3}$ & $2.51\cdot10^{-3}$ & $1.40\cdot10^{-2}$ & $1.08\cdot10^{-1}$ & $9.02\cdot10^{-3}$ & $2.70\cdot10^{-3}$ & $7.15\cdot10^{-2}$ & $1.49\cdot10^{-1}$ & $2.99\cdot10^{-2}$ & $5.91\cdot10^{-3}$ \\  
        LLY   & $1.07\cdot10^{-2}$ & $5.84\cdot10^{-2}$ & $5.40\cdot10^{-3}$ & $2.20\cdot10^{-3}$ & $1.18\cdot10^{-2}$ & $6.07\cdot10^{-2}$ & $5.40\cdot10^{-3}$ & $2.20\cdot10^{-3}$ & $4.09\cdot10^{-2}$ & $1.33\cdot10^{-1}$ & $5.40\cdot10^{-3}$ & $2.20\cdot10^{-3}$ \\  
        HD    & $3.40\cdot10^{-3}$ & $3.50\cdot10^{-2}$ & $5.40\cdot10^{-3}$ & $2.20\cdot10^{-3}$ & $3.80\cdot10^{-3}$ & $4.30\cdot10^{-2}$ & $5.40\cdot10^{-3}$ & $2.20\cdot10^{-3}$ & $1.00\cdot10^{-2}$ & $6.53\cdot10^{-2}$ & $5.40\cdot10^{-3}$ & $2.20\cdot10^{-3}$ \\  
        PG    & $9.20\cdot10^{-3}$ & $6.00\cdot10^{-2}$ & $7.20\cdot10^{-3}$ & $5.40\cdot10^{-3}$ & $1.03\cdot10^{-2}$ & $7.86\cdot10^{-2}$ & $2.03\cdot10^{-2}$ & $1.03\cdot10^{-2}$ & $1.52\cdot10^{-1}$ & $2.15\cdot10^{-1}$ & $5.35\cdot10^{-2}$ & $1.88\cdot10^{-2}$ \\ \hline  
    \end{tabular}
    \caption{Comparison of AdS, SABR and fSABR models.}
\end{table} 
\newgeometry{left=2cm, right=2cm, top=3cm, bottom=3cm}

\bibliographystyle{plain}
\bibliography{references}

\end{document}